\documentclass[12pt,preprint]{aastex}
\usepackage{amsmath}
\usepackage{mathrsfs}
\usepackage{latexsym,amsfonts,amssymb}
\usepackage{natbib}
\usepackage{graphicx}
\begin{document}
\title{Constraining the Lorentz Factor of a Relativistic Source from its Thermal Emission}

\author{Yuan-Chuan Zou\altaffilmark{1,2}, {K. S. Cheng}\altaffilmark{2}, F. Y. Wang\altaffilmark{2,3}}

\altaffiltext{1}{School of Physics, Huazhong University of Science and
Technology, Wuhan, 430074, China. Email: zouyc@hust.edu.cn (YCZ)}
\altaffiltext{2}{Department of Physics, University of Hong Kong, Hong Kong, China.}
\altaffiltext{3}{School of Astronomy and Space Science, Nanjing University, Nanjing 210093, China.}

\begin{abstract}
We propose a direct and simple method to measure the Lorentz factor of a relativistically expanding object such as gamma-ray bursts. Only three measurements, i.e. the thermal component of the emission, the distance and the variable time scale of the light curve, are used. When the uncertainties are considered, we will obtain a lower limit of the Lorentz factor instead. We apply this method to GRB 090618 and get a lower limit of 22 for the Lorentz factor. This method can be used for any relativistically moving object, such as gamma-ray bursts and soft gamma-ray repeaters.
\end{abstract}


\maketitle

\section{Introduction}
For a relativistically moving object it is crucial to know the
Lorentz factor ($\Gamma$). In the context of  astrophysical
objects, there are several phenomena which are claimed to be
associated with relativistic motions, such as gamma-ray bursts
(GRBs) \citep{2004RvMP...76.1143P}, active galactic nuclei (AGN)
\citep{1993ApJ...407...65G}, and soft gamma-ray repeaters (SGRs)
\citep{1991ApJ...366..240N}. Especially for GRBs, the Lorentz
factors of jets are believed to be about a few hundred.

The identification of high $\Gamma$ is from the compactness problem
\citep{1967MNRAS.135..345R}. There are several methods to constrain
the Lorentz factor of the relativistic motion, like superluminal
motion for the AGN \citep{1994ApJ...430..467V}, the afterglow
deceleration time of the GRBs \citep{1997ApJ...476..232M}, the
thermal emission from a standard fireball at the photosphere radius
\citep{2005ApJ...635..516N, 2007ApJ...664L...1P}, the compactness
problem for the observed high energy photons
\citep{1993A&AS...97...59F,2001ApJ...555..540L}, and the quiet period of the light
curves \citep{2010MNRAS.402.1854Z}. However, these methods are
either model-dependent (or parameter dependent)
\citep{1994ApJ...430..467V, 1997ApJ...476..232M,
2005ApJ...635..516N, 2007ApJ...664L...1P} or only upper/lower limits can be obtained
\citep{1993A&AS...97...59F, 2010MNRAS.402.1854Z}.

Our method can directly derive (or
constrain the lower limit of) $\Gamma$ if the thermal emission is identified,
while this thermal component has been observed in many cases,
such as GRB 090618 \citep{2011MNRAS.416.2078P}, GRB 090902B \citep{2012MNRAS.420..468P} and GRB 100724B \citep{2011ApJ...727L..33G}. \citet{Fan2012} suggested the thermal
component may even dominate the prompt emission in some GRBs.
We mainly concentrate on the GRBs, because the GRB jets are mostly believed
to be  pointing toward the observers. Considering the beaming effect of
the relativistic jet, spherical geometry can be safely taken only if the jet opening angle
$\theta_j > \frac{1}{\Gamma}$.
The general scenario of the GRB thermal emission is described by
\cite{2011MNRAS.415.1663T}: multiple conical shells are expanding, which
produce the thermal pulses, while the internal shocks from the collision of these shells produce the non-thermal emission at larger radii. The
superposition of these two kinds of emissions is the observed spectrum. Here we
concentrate on the thermal component to derive the Lorentz factor of
these shells. We describe the basic concept in section \S 2,
apply the method to GRBs in section \S 3, and discuss the results in
section \S 4.

\section{Basic concepts}
For a relativistically expanding shell, the thermal photons are
coupled with the electrons at the beginning. With the expansion of
the relativistic flow (either spherical or conical in shape), it
becomes optically thin at the photosphere, where the optical depth
drops down to unity. At this radius almost all the thermal photons
escape, their luminosity following the blackbody spectrum. Beyond this radius, as no more photons are produced, emission drops down quickly and can be neglected, the energy stored in the plasma will be
released by the internal or external shocks at larger radii, where
nonthermal emission (synchrotron radiation for instance) dominates.
Therefore, the observed thermal emission can be taken as blackbody
emission at the photosphere. \footnote{There is another possibility,
namely that in the optically thick stage, photons coupled with electrons are no longer produced. With the expansion of the relativistic shells, the shape of the spectrum becomes a Compton spectrum \citep{2013ApJ...764..143V}. The number density of the photons is then smaller than from a blackbody emission.
Consequently, the luminosity at the photosphere is lower than from a
blackbody with the same temperature. This leads to a lower limit for
the Lorentz factor. Though, in principle, this kind of thermal
emission can be distinguished from the blackbody emission (Plank
spectrum) by the spectral shape. It is not easy to distinguish
in real data. The combination leads
to a lower limit rather than a certain value, which is shown in
eq. (\ref{F_obs1}). Whether or not the Compton spectrum is being distinguished
from the Plank spectrum, eq. (\ref{F_obs1}) splits into either an
equality or a lower limit.}

Because of the beaming effect, for a blackbody sphere expanding with Lorentz factor $\Gamma$ ($\Gamma \gg 1$), the received flux is mainly emitted from the region within $\theta \le \frac{1}{\Gamma}$, where $\theta$ is the angle starting from the line of sight, as shown in Figure 1. The blackbody flux we can receive satisfies
\begin{equation}
F_{\rm thermal}^{\rm obs} \leq F_{\rm bb}^{} \simeq \frac{{\sigma [({1+z}){T_{\rm obs}}]^4 4 \pi (\frac{R_{\rm ph}}{\Gamma})^2 } }{{ 4 \pi D_{\rm L}^2}},
\label{F_obs1}
\end{equation}
where $F_{\rm thermal}^{\rm obs}$ is the observed flux of the thermal component, $F_{\rm bb}$ is the observed flux by assuming the radiation is from a blackbody in thermal equilibrium, $\sigma =5.67\times 10^{-5} {\rm erg\,cm^{-2}\,s^{-1}\,K^{-4}}$, $z$ is the redshift, $T_{\rm obs}$ is the observed temperature, $R_{\rm ph}$ is the photosphere radius of the expanding fireball, and $D_{\rm L}$ is luminosity distance. $F_{\rm thermal}^{\rm obs} \leq F_{\rm bb}^{}$ comes from the fact that the Thompson scattering photosphere can be larger than the absorption/radiation photosphere \citep[see][for more details]{2013ApJ...764..143V}. Notice that the equation is also applicable for a conical jet with opening angle $\theta_j > \frac{1}{\Gamma}$ when the line of sight is inside the jet angle.

\begin{figure}
\includegraphics[width=0.6\textwidth]{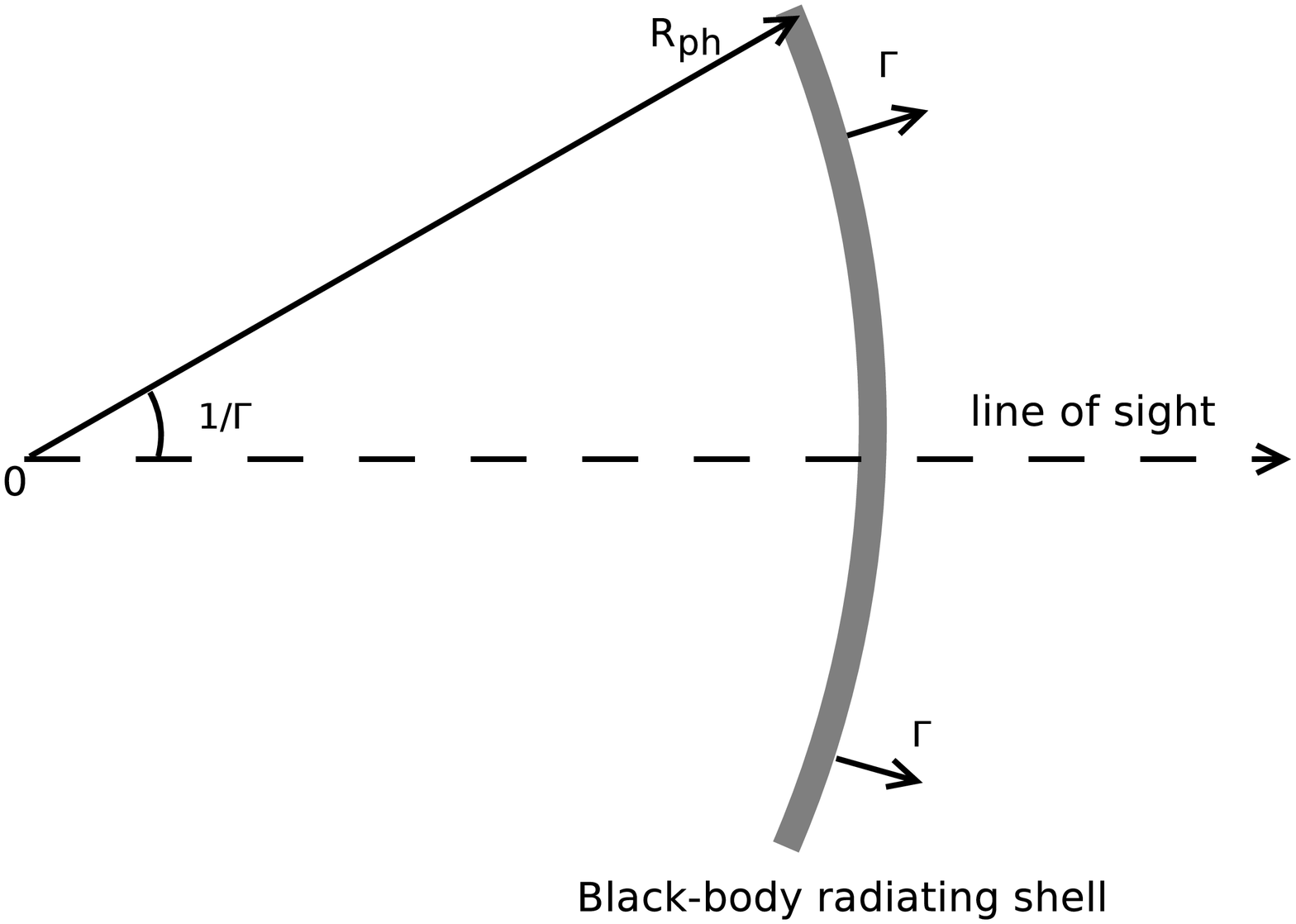}
\caption{Sketch for the conical radiating shell at the photosphere, which is moving with Lorentz factor $\Gamma$. Because of the beaming effect, roughly only the region inside $1/\Gamma$ can be seen by the observer. This determines the time scale of the thermal pulse.}
\end{figure}

As almost all the thermal photons in the plasma will be radiated at the photosphere suddenly, the observed duration of this emission is generally dominated by the angular time scale \citep{1999PhR...314..575P}, $\delta t_{\rm ang} \simeq \frac{(1+z)R_{\rm ph}}{2 \Gamma^2 c}$,
where $c$ is the speed of light.
In some cases the angular time scale is just a lower limit, e.g., for the thick shell case before the spreading radius, the observed time scale is dominated by the thickness of the shell, or several thermal pulses are too close to be distinguished, both cases imply  $\delta t_\oplus \geq \delta t_{\rm ang}$, i.e.
\begin{equation}
\delta t_\oplus \geq \frac{(1+z)R_{\rm ph}}{2 \Gamma^2 c}.
\label{t_ang}
\end{equation}
%On the other hand, if the electrons are being kept heated (e.g. the absorption opacity is greater than the scattering opacity), the radiating luminosity is smaller than from a blackbody [radiation though the spectrum being thermal????], because the photons are scattered into a higher energy by the electrons while no more photons can be produced\footnote{There is no possibility that the thermal-like radiation can be more luminous than a blackbody emission.}.
Combining eqs. (\ref{F_obs1}) and (\ref{t_ang}), we get the lower limit for the bulk Lorentz factor:
\begin{equation}
\Gamma \geq  \left[\frac{F_{\rm thermal}^{\rm obs} D_{\rm L}^2}{(1+z)^2 \sigma T_{\rm obs}^4 ({2  c \delta t_{\oplus}})^2}\right]^{\frac{1}{2}}.
\label{GammaBB2}
\end{equation}
As all the variables on the right hand side are observables, the $\Gamma$ is then a directly measurable quantity, by measuring the spectrum of the thermal component and the redshift of the object. The equality in the equation above applies in the ideal condition: the emitting region is geometrically thin, and is in thermal equilibrium.

There is another constraint from the fact that the thermal component is mainly released at the radius where the shell becomes optically thin, i.e., the optical depth $\tau=1$. This determines the radius of the photosphere \citep{2002ApJ...578..812M,2002MNRAS.336.1271D}
\begin{equation}
R_{\rm ph} = {{L \sigma_T}\over{8 \pi \Gamma^3 m_p c^3}},
 \label{Rph}
\end{equation}
where $L$ is the kinetic luminosity of the ejecta, which is taken as cold when the thermal photons are emitted, $\sigma_T$ is the Thompson cross-section and $m_p$ is the mass of the proton.
\cite{2007ApJ...664L...1P} considered this relation together with the blackbody radiation luminosity (i.e., eq. (\ref{F_obs1})) to derive the initial Lorentz factor without considering the time scale  (i.e., eq. (\ref{t_ang})).\footnote{Though they did not consider the angular time scale explicitly, when they argue that the thermal emission after the peak at around 8 s of GRB 970828 comes from the off-axis emission, the angular time scale shows the off-axis emission could not last a time of more than 10 s.
 In addition, if the followed thermal component was coming from the high latitude as \cite{2007ApJ...664L...1P} claimed, the temperature decay with time should be $t^{-1}$ but not $t^{-0.51}$ as observed. Therefore, we should conclude the decaying time scale of seconds is not from the off-axis emission.} One should consider the consistency between the three equations: eqs. (\ref{F_obs1}), (\ref{t_ang}) and (\ref{Rph}),
but keeping in mind the uncertainties of eq. (\ref{Rph}) arising from the facts: 1. the efficiency of conversion of the kinetic energy to the observed $\gamma$-ray energy is not known, it may include the Poynting flux making the efficiency even more uncertain; 2. the amount of electron-positron pairs is not known, which will affect the optical depth; 3. the matter should be fully ionized, otherwise another parameter characterizing the ionization should be introduced; 4. Eq. (\ref{Rph}) is only suitable when the Thompson scattering is the dominant cause of opacity; 5. $F_{\rm thermal}^{\rm obs} = F_{\rm bb}^{}$ is assumed. In the optimistic condition, one can combine eq. (\ref{Rph}) and eq. (\ref{F_obs1}) to get the Lorentz factor by introducing an efficiency parameter $Y\equiv L/L_\gamma$ ($L_\gamma$ is the observed $\gamma$-ray luminosity) \citep{2007ApJ...664L...1P}
\begin{equation}
\Gamma=\left[ \frac{\sigma \sigma_T (1+z)^2 Y L_\gamma T_{\rm obs}^2}{8\pi m_p c^3 D_L F_{\rm thermal}^{\rm obs \,{1/2}}} \right]^{1/4}.
\label{GammaBB3}
\end{equation}

\section{Application to GRBs}
%{\bf \it Gamma-ray bursts \\}
Taking the typical values of GRBs into eq. (\ref{GammaBB2}), one gets the scaling law:
\begin{equation}
\Gamma \geq  110 F_{\rm thermal, -6}^{\rm obs \,\frac{1}{2}} D_{\rm L, 28} (\frac{1+z}{2})^{-1} T_{\rm obs, 8}^{-2}  \delta t_{\oplus,-2}^{-1}.
\label{GammaGRB}
\end{equation}
The conventional notation $Q = Q_x \times 10^x$ is used throughout this letter.
The greatest uncertainty here is the choice of time scale $\delta t_\oplus \sim 0.01$ s, as the variability of the thermal component has not been identified in so small time bins \citep{2011MNRAS.416.2078P, 2013ApJ...770...32G, 2014arXiv1403.0374B}.
Though the observed typical time scale for the GRB pulses is of the order of 0.1 s \citep{2002MNRAS.330..920N}, they are believed to be from the internal shocks which occur farther out than the photospheric radius. Consequently, the time scale of the photosphere radius should be smaller. It may also be supported by the finding that the GRB pulses also contain millisecond variabilities \citep{2000ApJ...537..264W}.
%On the other hand, this choice is consistent with eq. (\ref{Rph}): $R_{\rm ph} = 1.85 \times 10^{12} L_{53} \Gamma_{2.5}^{-3}$ cm and $R_{\rm ph} = 2 \Gamma^2 c \delta t_\oplus$. For a conservative estimation, one may choose the lower limit $\Gamma_{\min} \simeq  11 F_{\rm thermal, -5}^{\rm obs \,\frac{1}{2}} D_{\rm L, 28} (\frac{1+z}{2})^{-1} T_{\rm obs, 8}^{-2}  \delta t_{\oplus,-1}^{-1}$.

There are several GRBs which were reported to have the thermal component in the prompt emission: GRB 060218 \citep{2006Natur.442.1008C}, GRB 090618 \citep{2011MNRAS.416.2078P}, GRB 090902B \citep{2012MNRAS.420..468P}, GRB 100316D \citep{2011MNRAS.411.2792S}, GRB 100724B \citep{2011ApJ...727L..33G}, GRB 120323A \citep{2013ApJ...770...32G} and  more in a recent analysis \citep{2014arXiv1403.0374B}. However, most of them have no light curve of the thermal component, which prevents us from estimating the time scale.
Among them, GRB 120323A is a strong short burst. It makes the time resolved spectral analysis possible, which shows a thermal time scale less than 0.1s \citep{2013ApJ...770...32G}. Unfortunately, the redshift of this burst is not known.

We take GRB 090618 as an example. It has been reported to have a thermal component, with $F_{\rm thermal}^{\rm obs} \simeq 10^{-6} {\rm erg\,cm^{-2}\,s^{-1}}$, $z=0.54$, $D_{\rm L} = 9.6\times 10^{27}$ cm, $kT_{\rm obs} \simeq 20$ keV \citep{2011MNRAS.416.2078P}. The most uncertainty is in the time scale of the thermal pulses, while $\Gamma$ is sensitive to it. We choose $\delta t \sim 0.01$s as a typical time scale for the thermal pulses.  Then the Lorentz factor is greater than $\sim 22$ by taking the values into eq. (\ref{GammaGRB}). The uncertainty of the time scale of the thermal component makes the constraint trivial. This constraint will be enhanced if the short time resolved spectral analysis can be performed, just like GRB 120323A.

We may consider the consistency of this method with other methods in the Lorentz factor constraint.
Taking the observed quantities into eq. (\ref{GammaBB3}), we get the Lorentz factor to be $130 Y^{1/4}$. This is consistent with the lower limit above (keep in mind, even if eq. (\ref{GammaBB3}) is not consistent with other methods, that might be because eq. (\ref{GammaBB3}) is not suitable for the reasons listed above).
One can also estimate the Lorentz factor by the afterglow light curve \citep{1997ApJ...476..232M}. From the X-ray afterglow of GRB 090618 no peak was observed, which indicates the peak time (deceleration time) should be smaller than 300 s, as shown in the light curve \citep{2010ApJ...708L.112D}. This upper limit of peak time (300 s) indicates a lower limit for the initial Lorentz factor of roughly $150$ obtained by using eq. (1) of \cite{2012ApJ...751...49L} taking observed quantities and assuming an ambient number density 1 ${\rm cm}^{-3}$. Hence the direct Lorentz factor is also consistent with the estimation from the afterglow deceleration time scale.
It should be consistent with the lower limit given by \citet{2001ApJ...555..540L}. However, the required high energy photons from GRB 090426 were not reported. On the other hand, a lower limit is always consistent with another lower limit obtained by a different method.

Besides the prompt emission, one can apply our method to estimate the Lorentz factor from the thermal emission of the afterglow. Consequently, the Lorentz factor characterizes the afterglow rather than the prompt emission region. GRB 090618 has also been reported with a thermal component from the afterglow at hundreds of seconds \citep{2011MNRAS.416.2078P}. However, it was in the steep decline stage \citep{2006ApJ...642..354Z}, which is mainly taken as indicating high latitude emission from the prompt radiating shell. Hence, this thermal component cannot be directly used to derive the Lorentz factor.

GRBs are widely known to be highly relativistic with $\Gamma \sim 100$ \citep{2004RvMP...76.1143P}. For a given $\Gamma \sim 100$, we can see from eq. (\ref{GammaGRB}), that the temperature $T_{\rm obs}$ must be relatively low (say roughly $< 10$keV), otherwise, the thermal radiation will be extremely high and cannot be missed by the spacecraft. The relatively low temperature might be the reason why so few GRBs have been firmly determined to have the thermal component.
% (only GRB 060218, 100316D and 090618 until 2010 \citep{2011MNRAS.416.2078P}, a few more were reported recently \citep{2013ApJ...770...32G, 2014arXiv1403.0374B}).

Notice that very few GRBs prompt emission are identified with a thermal component, and the thermal components are all identified with observed temperatures less than 10 keV. The main reason is that the current on-boading satellite missions mainly cover relatively high energy bands (more than tens of keV). It may be resolved by the lower band soft $\gamma$-ray telescopes scheduled to be launched, like SVOM \citep{2012MSAIS..21..162G}, JANUS \citep{2009SPIE.7435E...5F} and LOFT \citep{2015arXiv150102772A}. After identifying the thermal component, one needs to identify the time scales of the thermal pulses, which means the time bin should be small enough to perform a spectral analysis. This is still challenging but might be available for strong soft bursts in the near future. Otherwise, only lower limits can be achieved.

%The method can also be applied to the early X-ray afterglow in the steep decay phase, which is taken to be the high latitude emission from the prompt shell.

\section{Discussion}

To clarify, we emphasize the scenario for the thermal and non-thermal components in the following. The relativistic warm shells produced from the central engine are ejected stochastically. With the shells expanding spherically the optical depth for the trapped thermal photons decreases. At the transparent radius, the shell emits almost all the thermal photons, which are observed as the thermal component. Each shell will produce one thermal pulse with a short time scale. The combination of any two shells produces one internal shock, which generally occurs at a larger radius, and produces the non-thermal component. This non-thermal component has a longer time scale. The superposition of these two components is what we observe. For the purpose of obtaining the Lorentz factor, we need to extract the thermal component.
For GRBs, the thermal component should never be dominant. Otherwise, there will be more energy left in the plasma. This is related to the energy budget problem and the efficiency problem \citep{2004RvMP...76.1143P}. This makes the extraction of the thermal light curve more difficult.

In general, this method can be applied to any object radiating thermal emission, and with a relativistic velocity. There are several other objects to which this method may be potentially applicable, though shortcomings are obvious at the present stage.
SGRs are also thought to be relativistic events \citep[for example]{2003ApJ...593L..85C}. However, the blackbody emission in this case is mainly taken to be generated from the progenitor \citep{2005Natur.434.1098H}.
This method cannot be directly applied to AGN, as AGN jets are mainly pointing to other directions, they are all continuous jets, and there is no confirmed thermal radiation \citep{1998MNRAS.299..433F}.
X-ray binaries are believed to have relativistic jets too. However, the number of on-axis case is very limited, and the thermal component is not easily distinguished as arising from the jet or from the disk.
Tidal disruption events also have mildly relativistic ejecta. But no associated thermal component has been identified. The low Lorentz factor may be beyond the restriction $1/\Gamma \leq \theta_j$, and is not easy to obtain by our method.
This method may also be applied to the GRB afterglow stage if there is a thermal component.

We should also notice that it is applicable only if the following conditions are satisfied: 1. The jet opening angle should be greater than $1/\Gamma$, otherwise the shortest time scale is not $R_{ph}/2\Gamma^2 c$. Fortunately, for the GRBs the jet opening angle is generally greater than $1/\Gamma$. 2. The line of sight should be inside the jet opening angle, and this is always the case for GRBs.
3. Though in the GRB fireball scenario, it is unlikely the ejecta is still accelerating at the photosphere radius, it is possible in general. Then the derived Lorentz factor is a lower limit again.

One can also constrain the Lorentz factor by other methods, especially the upper limits. By combining several methods, once the Lorentz factor is constrained in a very narrow range without inconsistency, that means the equality of  eq. (\ref{GammaBB2}) can be applied. This indicates that for such a burst the thermal emission comes from a thin shell and no heating of the electrons occurred.

We thank Prof. Kevin MacKeown for a critical reading of the manuscript. YCZ acknowledges helpful discussions with Tsvi Piran, Zigao Dai, Bing Zhang, Xuefeng Wu, Xiwei Liu, Weihua Lei, Rongfeng Shen and Rodolfo Barniol Duran. This work is supported by the Chinese-Israeli Joint Research Project (Grant No. 11361140349), the National Basic Research Program of China (973 Program, Grant No. 2014CB845800), and the National Natural Science Foundation of China (Grants No. U1231101). KSC is supported by the CRT Grants of the Government of the Hong Kong SAR under HKUST4/CRF/13G.

\iffalse
Prove: $[\hat{p}_x^2, f(x)]=-\hbar^2 \frac{\partial ^2 f}{\partial x^2} - 2 i \hbar \frac{\partial f}{\partial x} \hat{p}_x$\\
\begin{eqnarray*}
[\hat{p}_x^2, f(x)] \Psi
&=& \hat{p}_x^2 (f(x) \Psi) - f(x) (\hat{p}_x^2  \Psi) \\
&=& -i \hbar \frac{\partial }{\partial x} \left[ -i \hbar \frac{\partial }{\partial x} (f \Psi) \right]- f(x) (\hat{p}_x^2  \Psi) \\
&=& -i \hbar \frac{\partial }{\partial x} \left[ -i \hbar f \frac{\partial }{\partial x}  \Psi -i \hbar \Psi \frac{\partial }{\partial x} f  \right]- f(x) (\hat{p}_x^2  \Psi) \\
&\equiv& -i \hbar \frac{\partial }{\partial x} \left[ -i \hbar f  \Psi_x -i \hbar \Psi f_x \right]- f(x) (\hat{p}_x^2  \Psi) \\
&=&  \left[- \hbar^2 f_x  \Psi_x - \hbar^2 f \Psi_{xx} - \hbar^2 \Psi_x f_x -\hbar^2 f_{xx} \Psi   \right]- f(x) (\hat{p}_x^2  \Psi) \\
&=&  \left[- \hbar^2 f_x  \Psi_x  - \hbar^2 \Psi_x f_x -\hbar^2 f_{xx} \Psi   \right]\\
&=& - 2 i \hbar \frac{\partial f}{\partial x} \hat{p}_x \Psi -\hbar^2 \frac{\partial ^2 f}{\partial x^2} \Psi .
\end{eqnarray*}
\fi

\end{document}